\documentclass[numberedappendix]{emulateapj}
\usepackage{apjfonts}
\usepackage{hyperref}
\usepackage{todonotes}
\newcommand{\dof}{{\rm dof}}
\shorttitle{Sub-mm QPOs in Sgr A* accretion flow}
\shortauthors{Shcherbakov \& McKinney}

\begin{document}
\title{SUBMILLIMETER QUASI-PERIODIC OSCILLATIONS \\IN MAGNETICALLY CHOKED ACCRETION FLOW MODELS OF SgrA*}
\author{Roman V. Shcherbakov\altaffilmark{1,2,3}, Jonathan C. McKinney\altaffilmark{2,4}}
\email{roman@astro.umd.edu}
\altaffiltext{1}{\url{http://astroman.org} \linebreak Department of Astronomy, University of Maryland, College Park, MD 20742, USA}
\altaffiltext{2}{Joint Space Science Institute, University of Maryland, College Park MD 20742, USA}
\altaffiltext{3}{Hubble Fellow}
\altaffiltext{4}{Physics Department, University of Maryland, College Park, MD 20742-4111, USA}

\begin{abstract}
High-frequency quasi-periodic oscillations (QPOs) appear in general-relativistic magnetohydrodynamic simulations of magnetically choked accretion flows around rapidly rotating black holes (BHs).
We perform polarized radiative transfer calculations with ASTRORAY code to explore the manifestations of these QPOs for SgrA*.
We construct a simulation-based model of a radiatively inefficient accretion flow and find model parameters by fitting the mean polarized source spectrum.
The simulated QPOs have a total sub-mm flux amplitude up to $5\%$ and a linearly polarized flux amplitude up to $2\%$. The oscillations reach high levels of significance $10-30\sigma$ and high quality factors
$Q\approx5$. The oscillation period $T\approx100M\approx35$~min corresponds to the rotation period of the BH magnetosphere that produces a trailing spiral in resolved disk images.
The total flux signal is significant over noise for all tested frequencies $87$~GHz, $230$~GHz, and $857$~GHz and inclination angles $10^\circ,$ $37^\circ,$ and $80^\circ$.
The non-detection in the $230$~GHz Sub-Millimeter Array light curve is consistent with a low signal level and a low sampling rate.
The presence of sub-mm QPOs in SgrA* will be better tested with the Atacama Large Millimeter Array.
\end{abstract}


\section{INTRODUCTION}
Quasi-periodic oscillations (QPOs) in the emission from black hole (BH) accretion disks and jets
are found in systems with both stellar mass BHs \citep{Remillard:2006} and supermassive BHs (SMBHs) \citep{Gierlinski:2008wj,Reis:2012sc}.
The high-frequency QPOs (HFQPOs) with a period $T$ about the orbital period at the innermost stable circular orbit (ISCO)
potentially probe the region close to the event horizon, offering a chance to test accretion and jet theories in the strong gravity regime.
There have been multiple claims of the HFQPOs from the SMBH SgrA* in the Milky Way center, which then provides a unique opportunity to study the HFQPOs up-close.
Henceforth, we set the speed of light and gravitational constant to unity ($c=1$ and $G=1$), such that $1M=21$~s for SgrA*
with the BH mass $M_{\rm BH}=4.3\times10^{6}M_\odot$ \citep{Ghez2008,Gillessen:2009oo}.

A HFQPO period commonly claimed for SgrA* is $T=17{\rm min}=48M$. This was suggested in the K band during a flare \citep{Genzel:2003aj},
in $7$~mm data by \citet{Yusef-Zadeh:2011os}, and in images obtained with very long baseline interferometry (VLBI) by \citet{Miyoshi:2011ah}.
Periods of $T=28{\rm min}=79M$ \citep{Genzel:2003aj},  $T=23{\rm min}=65M,$ and $T=45{\rm min}=127M$ \citep{Trippe:2007oa,Hamaus:2009fs} were reported in the IR observations of flares.
However, \citet{Do:2009lk} analyzed a long K band light curve and found no statistically significant power spectrum density excess.
A statistically significant longer period of $T=2.5-3{\rm hrs}\approx470M$ was reported by \citet{Mauerhan:2005rh} in $3$~mm data.
\citet{Miyoshi:2011ah} claimed a range of periods from $17$~min to $56$~min due to the spirals with multiple arms, but did not compute significance.

Physical origins of the HFQPOs are highly debated. Models are often based on an ISCO orbital frequency,
an epicyclic frequency, and frequencies of various pressure and gravity modes (e.g., \citealt{Kato:2001aj,Kato:2004fw,Remillard:2006,Wagoner:2008pq}).
The underlying physics involves beat oscillations \citep{Klis:2000he}, resonances between flow modes and normal frequencies in general relativity (GR)
\citep{Abramowicz:2001ja}, trapped oscillations \citep{Nowak:1991jp}, parametric resonances \citep{Abramowicz:2003lu}, and disk magnetospheric oscillations \citep{Li:2004qp}.

The analytic methods were followed by blind QPO searches in magnetohydrodynamic (MHD) simulations.
Non-GR 2D MHD simulations exhibited spiral patterns of Rossby waves detectable in simulated light curves \citep{Tagger:2006jw,Falanga:2007uj}.
Non-GR 3D MHD simulations of thick accretion disks by \citet{Chan:2009ja} developed the QPOs with a period $T=39M$
\footnote{We rescale reported periods to the current value of SgrA* BH mass.} in a simulated X-ray light curve.
The 3D GRMHD simulations and radiative transfer by \citet{Schnittman:2006ia} revealed weak transient QPOs.
Similar simulations and radiative transfer by \citet{Dolence:2009wb,Dolence:2012xs} showed a spiral structure
producing oscillations with periods $T=6-9{\rm min}=17-25M$ in simulated NIR and X-ray light curves.
The GRMHD simulations of tilted disks produce tentative QPOs in dynamical quantities with $T\approx170M=1$~hr \citep{Henisey:2012la}.

Such MHD simulations start with a weak magnetic field, which is amplified by the magneto-rotational instability (MRI) that generates incoherent turbulence.
However, when magnetized gas falls onto a BH, the disk becomes saturated with more magnetic flux than the MRI can generate \citep{McKinney2012}.
The 3D GRMHD simulations of radiatively inefficient accretion flows (RIAFs, as applicable to SgrA*) with ordered magnetic flux were performed by \citet{McKinney2012}.
The resultant magnetically choked accretion flow (MCAF) has a BH magnetosphere that significantly affects the sub-Keplerian equatorial inflow.
The simulations showed high-quality disk-BH magnetospheric QPOs in dynamical quantities with an $m=1$ (one-arm) toroidal mode and a rotating inflow pattern in the equatorial plane.

We quantify the QPO signal and its statistical significance in simulated SgrA* light curves based on the GRMHD simulations of MCAFs.
We do a targeted search for the known QPO period $T\approx100M$. In Section~\ref{sec:GRMHD} we describe the 3D GRMHD simulations and the application to SgrA*.
We perform GR polarized radiative transfer calculations with ASTRORAY code, fit the SgrA* mean polarized spectrum, and find the best-fitting model parameters.
In Section~\ref{sec:timing} we describe timing analysis. We study the light curves of the best-fitting model viewed at different inclination angles $\theta$.
We find statistically significant QPOs in total and some linearly polarized (LP) fluxes.
We image a correspondent equatorial plane spiral wave.
In Section~\ref{sec:discussion} we compare our simulated QPOs with previous work and discuss the observability in SgrA*.

\section{SgrA* MODEL BASED ON GRMHD SIMULATIONS}\label{sec:GRMHD}
\subsection{GRMHD Simulations}
The initial gas reservoir is a hydrostatic torus \citep{Gammie:2003in}, within which magnetic field loops are inserted.
The MRI action on the initial field leads to MHD turbulent accretion that eventually causes magnetic flux to saturate near the BH \citep{McKinney2012}.
We focus on a simulation with a dimensionless spin $a_*=0.9375$, which is close to $a_*\approx0.9$ favored in simulation-based modeling of the SgrA*
spectrum and the emitting region size \citep{Moscibrodzka:2009,Dexter:2010lk,Shcherbakov:2012appl}.
The simulation is performed in spherical coordinates $(r,\theta,\phi)$ with resolution $N_r\times N_\theta\times N_\phi=272 \times 128\times 256$.
It reached a quasi-steady state by time $t=8,000M$ and ran till $t=28,000M$.
In steady state near the BH event horizon, the sub-Keplerian inflow is balanced
against the BH magnetosphere resulting in vertical compression of the disk.

The BH magnetosphere and disk exhibit the QPOs in dynamical quantities such as the magnetic field energy density.
A toroidal wobbling mode with $m=1$ is eminent in the jet polar region and disk plane.
It was identified with pattern rotation of the BH magnetospheric region pierced by the infalling matter streams.
The streams form due to magnetic Rayleigh-Taylor instabilities (e.g. \citealt{Stone:2007qw}).
The pattern rotates with an angular frequency $\Omega_F\approx0.2\Omega_H$, where $\Omega_H=a_*/(2r_H)$ is the BH angular frequency and
$r_H=(1+\sqrt{1-a_*^2})M$ is the horizon radius. The angular frequency $\Omega_F$ is close to the rotation frequency $\approx0.27\Omega_H$ of the field lines attached
to the BH at the equatorial plane in a paraboloidal magnetospheric solution \citep{Blandford1977}.

\subsection{SgrA* Accretion Flow Model}
We use this MCAF 3D GRMHD simulation to model the SgrA* accretion flow.
We follow \citet{Shcherbakov:2012appl} to define the electron temperature and extrapolate quantities to outer radii $r>50M$.
A power-law extension of density to $r>50M$ is $n\propto r^{-\beta}$, while the proton temperature is continued as $T_p\propto r^{-1}$.
The magnetic field strength is extended as $b\propto \sqrt{n T_p}\propto r^{(-1-\beta)/2}$ to preserve a constant local ratio of magnetic field energy to thermal energy.
The slope $\beta$ is found by connecting the known density at $r=3\times10^5M$ to the density in the inner region \citep{Shcherbakov:2010cond}.
Correct simultaneous evolution of the simulations and the radiation field is considered, despite the radiative transfer is conducted in post-processing.

We focus on the accretion disk as the source of SgrA* emission and will consider jet emission (e.g., \citealt{Falcke:2004oq}) in future studies.
The simulated matter density is artificial near the polar axis, because matter is injected there to avoid
an exceedingly high local ratio of magnetic energy to rest-mass energy that is difficult for GRMHD codes to evolve.
The injected material does not change flow dynamics because it is energetically negligible.
Nevertheless, a small amount of hot matter in the polar region can shine brightly as revealed by \citet{Moscibrodzka:2009}.
The matter densities are zeroed out in a bipolar cone with an opening angle $\theta=26^\circ$.
If that artificial matter was not removed, then none of our models would be consistent with the observed image size at $230$~GHz \citep{Doeleman:2008af} and
the observed polarized SgrA* spectrum.

The radiative transfer is performed with our ASTRORAY code \citep{Shcherbakov:2011inter,Shcherbakov:2012appl}.
We compute radiation over a quasi-steady simulation period between $t=8,000M$ and $t=28,000M$.
Following the previous work, we fit the total flux of SgrA* at $87-857$~GHz, the LP fraction at $87$~GHz, $230$~GHz, and $345$~GHz,
and the circular polarization (CP) fraction at $230$~GHz and $345$~GHz. We vary the heating constant $C$, which determines the electron temperature $T_e$ close to the BH,
the accretion rate $\dot{M}$, and the inclination angle $\theta$.
Fitting the mean SgrA* spectrum with the mean simulated spectrum we reach $\chi^2/\dof=1.55$ for $\dof=9$,
which is a better agreement than in our prior work based on weakly magnetized simulations \citep{Penna:2010dj,Shcherbakov:2012appl}.
The correspondent values of parameters are $T_e=3.2\times10^{10}$~K at $6M$ distance from the center, $\dot{M}=1.0\times10^{-8}M_\odot{\rm yr}^{-1}$, and $\theta=37^\circ$.
We then perform a timing analysis of the light curves from a number of models.

\section{TIMING ANALYSIS}\label{sec:timing}
\subsection{Oscillations in Light Curves and Images}
\begin{figure}[htbp]
    \centering\plotone{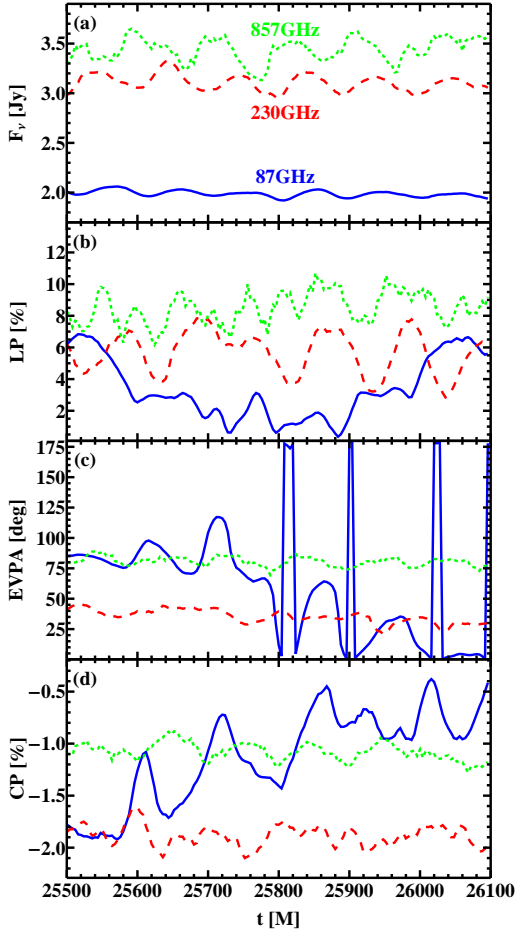}
    \caption{Polarized light curve fragments for the best-fitting model with the inclination angle $\theta=37^\circ$ at the optically thick $87$~GHz (blue solid line),
at $230$~GHz (red dashed line) with the optical depth about unity, and at the optically thin $857$~GHz (green dotted line).}
    \label{fig:QPOfreq}
\end{figure}
Let us first demonstrate the oscillations in the light curves. In Figure~\ref{fig:QPOfreq} we show the light curves at times $t=25,500M-26,100M$
for the best-fitting model with the inclination angle $\theta=37^\circ$.
The light curves are computed for three frequencies with different optical depth: radiation at $87$~GHz is optically thick, the optical depth at $230$~GHz is about $\tau\sim1$, while
radiation is optically thin at $857$~GHz.  The total flux (top panel) shows regular oscillations with the amplitude $\Delta F\approx0.05$~Jy at $87$~GHz
and $\Delta F\approx0.15$~Jy at $230$~GHz.  Fluctuations at $857$~GHz with the amplitude $\Delta F\approx0.2$~Jy are less regular.
The LP fraction fluctuates at $2\%$ level at all three frequencies, which translates into the relative variations of up to $50\%$ and the absolute LP flux variations $\Delta F\approx 0.06$~Jy.
The LP and CP fractions and the electric vector position angle (EVPA) exhibit substantial variations over long timescales at $87$~GHz.
The variations of the EVPA at $230$~GHz and $857$~GHz are about $5^\circ-10^\circ$. The CP fraction oscillates by $0.3\%$ at $87$~GHz and by $0.15\%$ at the higher frequencies.
The absolute CP flux variations are $\Delta F\approx0.005$~Jy at $230$~GHz.

\begin{figure}[htbp]
    \centering\plotone{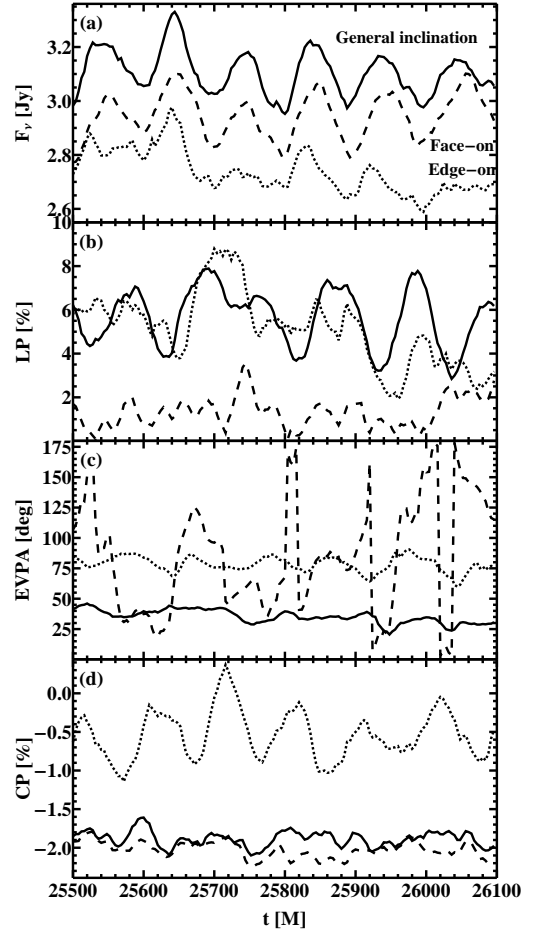}
    \caption{Polarized light curve fragments at $230$~GHz for the best-fitting $T_e$ and $\dot{M}$ for several inclination angles $\theta$:
best-fitting $\theta=37^\circ$ (solid line),  face-on $\theta=10^\circ$ (dashed line), and edge-on $\theta=80^\circ$ (dotted line).}
    \label{fig:QPOcurve}
\end{figure}
In Figure~\ref{fig:QPOcurve} we show how the amplitude of oscillations depends on the inclination angle $\theta$ at $230$~GHz.
Shown are the light curves for the best-fitting inclination angle $\theta=37^\circ$, for almost face-on $\theta=10^\circ$,
and for almost edge-on $\theta=80^\circ$.  The total flux exhibits the same oscillation amplitude $\Delta F\approx0.15$~Jy independent of $\theta$.
The edge-on and the best-fitting cases produce comparable variations of the LP fraction, while cancelations of the polarized fluxes emitted across the flow
lower both the mean and the fluctuation amplitude of the LP fraction in the face-on case. Correspondingly, the EVPA fluctuates dramatically in the face-on case.
The CP fraction oscillates at $0.5\%$ level in the edge-on case, while the other cases exhibit oscillation amplitude $0.15\%$.

\begin{figure*}[htbp]
    \centering\plotone{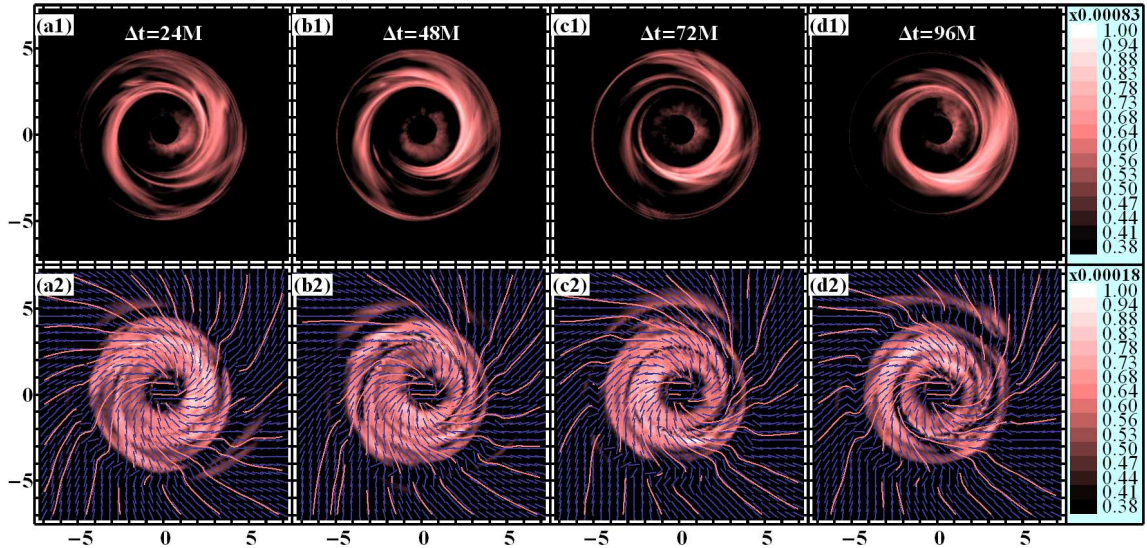}
    \caption{Images of a face-on disk at $230$~GHz: the total intensity images (top row) and the LP intensity images (bottom row).
    Strokes indicate the EVPA direction. The rotating spiral pattern is clearly visible.}
    \label{fig:images}
\end{figure*}
The face-on accretion flow images with $\theta=10^\circ$ are shown in Figure~\ref{fig:images}: the time series of the total intensity images in the top row and of the LP intensity images in the bottom row.
The total intensity images show a clear one-arm spiral rotating with a period $T\approx100M$.
The LP intensity spiral is spatially offset from the total intensity spiral, as the region of the brightest total intensity exhibits the strongest LP cancelations.
The total intensity spiral looks similar to that in \citet{Dolence:2012xs}, despite different angular velocities.
Note that we report the intensity images, while \citet{Dolence:2012xs} showed the images of the dynamical quantities.

\subsection{Statistical Analysis}
Let us quantify significance of the QPOs. Following \citet{Papadakis93} we start with an autocovariance
\begin{equation}\label{eq:autocovar}
\hat{R}(k)=\frac{1}{N}\sum^{N \Delta t-|k|}_{t=1\Delta t}(x_t-\bar{x})(x_{t+|k|}-\bar{x}),
\end{equation} where $k=0,\pm1\Delta t,...,\pm(N-1)\Delta t$ and $x_t$ is the sample of simulated fluxes normalized to have its mean $\bar{x}$ equal unity.
Then we compute a periodogram
\begin{equation}
I(R)=\frac{\Delta t}{2\pi}\sum^{(N-1) \Delta t}_{k=-(N-1) \Delta t}\hat{R}(k)\cos \omega k, \quad -\frac{\pi}{\Delta t}\le \omega \le\frac{\pi}{\Delta t}.
\end{equation}
In our analysis $\Delta t=4M$, which appears large enough to avoid aliasing at periods $T>50M$.
All periodograms  are log-smoothed \citet{Papadakis93} to $0.08$dex as a compromise between stronger random noise and larger smearing of the QPO peaks.

The determination of statistical significance of the QPOs involves comparison of the simulated periodogram with the random noise periodograms.
We follow the procedure in \citet{Timmer:1995wa} for random noise generation. We employ a log-smoothed to $3.0$dex periodogram
of the simulated light curve as the underlying non-QPO periodogram. This approximation produces a smooth curve comparable to fits of the non-QPO power spectrum with a power-law of a broken power-law.
We draw the random noise Fourier transform from a normalized Gaussian distribution and perform the inverse Fourier transform to generate the noise light curves.
We then compute noise autocovariances and periodograms. We find a $3\sigma$ significance curve based on $2,592$ random noise samples.
We do not correct for the blind search, since the periodograms are binned, and we target the QPOs with a period $T\approx100M$.
\begin{figure*}[htbp]
    \epsscale{0.85}\plotone{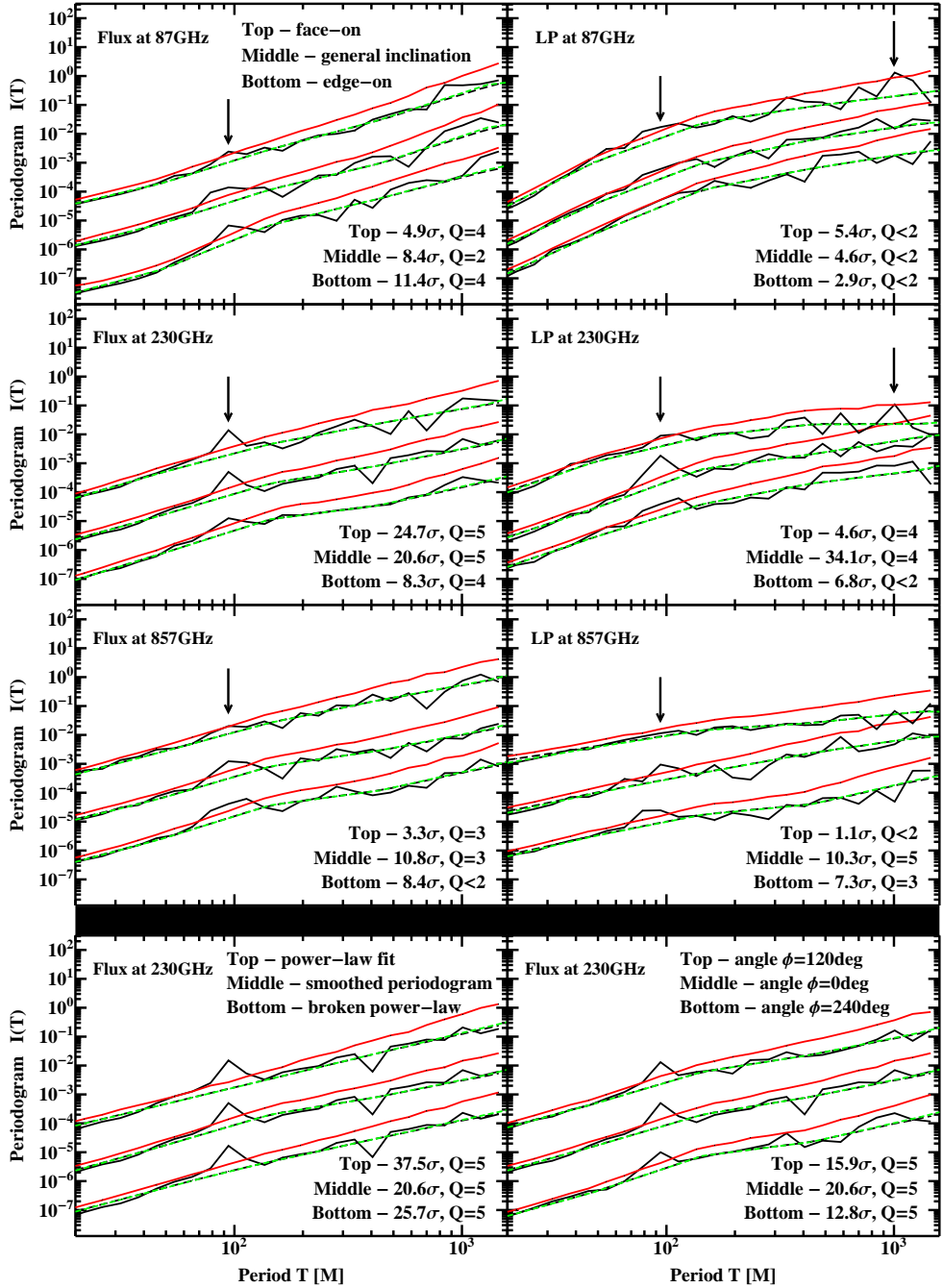}
    \caption{Periodograms of the total flux and the LP fraction light curves for the frequencies $87$~GHz, $230$~GHz, and $857$~GHz and
the inclination angles $\theta=10^\circ$ (top curves on top six panels), $\theta=37^\circ$ (middles curves), and $\theta=80^\circ$ (bottom curves),
while $T_e$ and $\dot{M}$ are fixed at their best-fitting values: simulated periodograms (black/dark solid lines), $3\sigma$ significance curves (red/light solid lines),
log-smoothed to $3.0$dex source periodograms (black/dark dashed lines), and geometric means of random noise periodograms (green/light dashed lines). The top six panels employ the log-smoothed
to $3.0$dex source periodogram as the underlying non-QPO periodogram and the azimuthal viewing angle $\phi=0\rm deg$.
The bottom left panel shows the results for the different approximations of the non-QPO periodogram.
The bottom right panel shows the results for the different azimuthal viewing angles.}
    \label{fig:periodogram}
\end{figure*}

The periodograms with their $3\sigma$ significance curves are depicted in the top six panels of Figure~\ref{fig:periodogram} for several frequencies and inclination angles.
The top curves in each panel are for almost face-on inclination $\theta=10^\circ$, the middle curves are for $\theta=37^\circ$,
and the bottom curves are for almost edge-on $\theta=80^\circ$.  We report the statistical significance levels and quality factors $Q=T_{\rm QPO}/{\rm FWHM}$.
The total flux and the LP fraction periodograms exhibit peaks significant to $5-30\sigma$ with $Q=2-5$ at the period $T\approx100M$ for most studied frequencies and inclination angles.
Our method allows for the maximum measurable quality of $Q=8-11$. A face-on disk shows weaker QPOs, which are not significant at $857$~GHz.
The LP fraction oscillations are weak for a face-on disk for all frequencies due to random cancelations of the LP.
The total flux and the LP fraction at $230$~GHz show the strongest oscillations, which further encourages SgrA* observations at $1.3$~mm wavelength.
We also detect marginally significant LP fraction oscillations with a period $T=1000M\approx4$~hr.
The bottom left panel of Figure~\ref{fig:periodogram} shows the analysis for the different approximations to the non-QPO periodogram: the power-law fit, the broken power-law fit, and
the log-smoothed to $3.0$dex source periodogram. The bottom right panel of Figure~\ref{fig:periodogram} shows the analysis for the different azimuthal viewing angles $\phi=0\rm deg$,
$\phi=120\rm deg$, and $\phi=240\rm deg$. The QPO peaks in these six cases stay prominent despite the significance level (number of sigmas) varies by $50\%$.
The significant QPO peaks among the cases presented in Figure~\ref{fig:periodogram} (top panels) stay significant, when we switch to the broken power-law fits to the non-QPO periodograms.

\begin{figure}[htbp]
    \centering\plotone{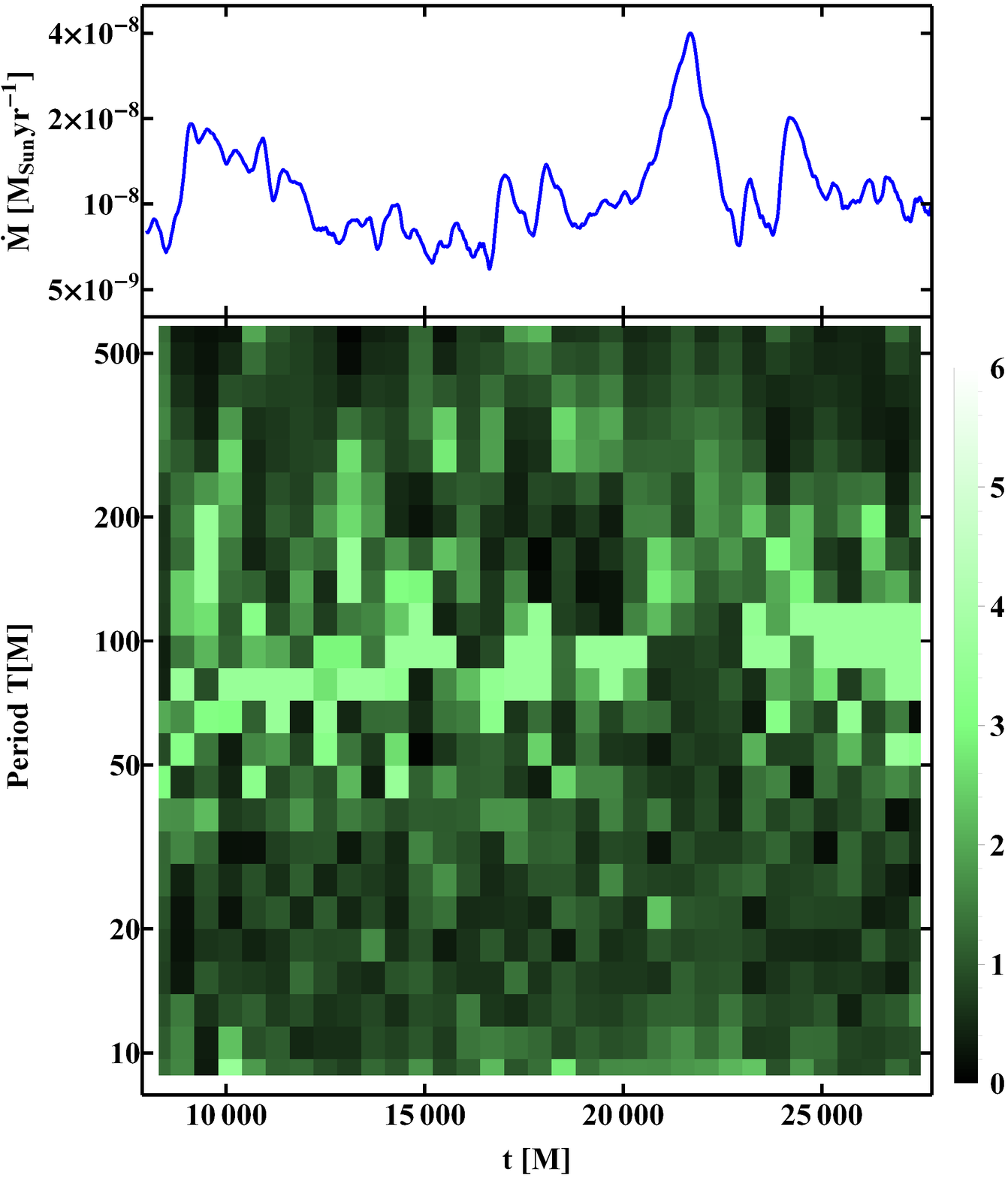}
    \caption{Accretion rate dependence on time $\dot{M}(t)$ (top) and normalized spectrogram of the total flux light curve (bottom) for the simulation intervals with $\Delta t=600M$.
The normalized spectrogram shows ratios of the log-smoothed to $0.08$dex periodograms over the log-smoothed to $3.0$dex periodograms. The higher ratio and the lighter color indicate the QPOs.}
    \label{fig:spectrogram}
\end{figure}
We characterize presence of the oscillations and stability of the oscillation period by a spectrogram in Figure~\ref{fig:spectrogram}.
The spectrogram indicates that most of the time oscillations with a period $T=90-100M$ are present.
However, no oscillations occur around time $t=22,000M$, when the accretion rate peaks due to weaker magnetic field.

\section{DISCUSSION AND CONCLUSIONS}\label{sec:discussion}
\subsection{Summary and Comparison to Previous Work}
Here we report the QPOs in the simulated SgrA* light curves for models based on the state-of-the-art 3D GRMHD simulations of the magnetically choked RIAFs.
The minimization procedure produces a fit with $\chi^2/\dof=1.55$ for $\dof=9$ to the mean polarized sub-mm source spectrum.
The correspondent simulated total flux light curve shows regular oscillations with the period $T\approx100M\approx35$~min and the amplitude $\Delta F\approx0.15$~Jy at $230$~GHz.
Less regular fluctuations with $\Delta F\approx0.2$~Jy are seen at $857$~GHz.
Weaker oscillations with $\Delta F\approx0.05$~Jy are seen at $87$~GHz, which probes the optically thick emission from $\sim10M$ radius.
The LP fraction exhibits periodic modulations at $50\%$ relative level, but the absolute LP flux amplitude is only about $\Delta F_{LP}\approx 0.06$~Jy.
The QPOs are significant above $3\sigma$ in the total flux light curves for all tested inclination angles $10^\circ,$
$37^\circ,$ and $80^\circ$ and frequencies $87$~GHz, $230$~GHz, and $857$~GHz, while the LP fraction shows less prominent QPOs at $87$~GHz and in a face-on case.

Our main $T\approx35$~min period is longer than the claimed SgrA* observed period $17-20$~min, while the simulated periods $T=6-9$~min in \citet{Dolence:2012xs} are shorter.
Their simulation has the same BH spin $a_*=0.9375$ as does our simulation, but the simulation by \citet{Dolence:2012xs} reaches a relatively weak BH horizon magnetic flux
and produces a thinner disk with height-to-radius ratio of $H/R\sim 0.2$. The resultant MRI-dominated accretion flow has a Keplerian rotation and shorter QPO periods.
They did not identify their QPO mechanism, although they noted their turbulence is unresolved \citep{Shiokawa:2012iq} and this might lead to artificial QPOs \citep{Henisey:2009gk}.
Our MCAF model has a sub-Keplerian rotation and the QPOs driven by the interaction of the disk with the rotating BH magnetosphere \citep{Li:2004qp} that leads to longer periods.
Our relatively thick disk with $H/R\sim 0.6$ is expected for a RIAF, and our simulations resolve well the disk turbulence \citep{McKinney2012} suggesting the QPOs are robust.
Based upon these works, SgrA* QPOs might be explained by $a_*=0.9375$ with an intermediate gas rotation rate, magnetization, or $H/R$.

Flow cooling, whose marginal importance for SgrA* was suggested by \citet{Drappeau:2012dq},
can self-consistently choose the disk thickness $H/R$ in simulations. In MCAFs, the steady-state BH horizon magnetic
flux has a positive correlation with $H/R$ \citep{McKinney2012}, so cooling can lead to more Keplerian rotation and a weaker magnetosphere,
and then the QPO period from MCAFs could be comparable to claimed for SgrA*.

\subsection{Observing QPOs in SgrA*}\label{subsec:observations}
We showed that the QPOs, though highly significant, have a maximum sub-mm amplitude of $5\%$ or $\Delta F\sim0.15$~Jy.
Low sensitivity and low sampling rate of current sub-mm instruments might prohibit observational detection of such oscillations \citep{Marrone:2006phd}.
The SMA achieves $5\%$ accuracy and samples every $10$~min at $1.3$~mm with a correspondent $20$~min Nyquist period \citep{Marrone:2008ep}.
A weak signal with $T\sim30$~min period is readily masked by aliasing and noise in the SMA data.
The total flux QPO amplitude $\Delta F\approx0.15$~Jy is larger than the LP flux amplitude $\Delta F\approx0.06$~Jy. However, the observational error of the total flux can also be larger.
The SMA measures the LP flux to the leakage level of $\sim0.3\%=9$~mJy, while the total flux is measured to $\sim0.7\%=20$~mJy due to calibration uncertainties \citep{Marrone:2007}.
Then it is about equally difficult to detect the total flux oscillations and the LP flux oscillations.

The ALMA gives more hope in detecting SgrA* sub-mm QPOs. It covers a wide frequency range $84-720$~GHz, has a collecting area of $\sim7\times10^3 {\rm m}^2$
about $30$ times that of the SMA, and can sample every few minutes \citep{Brown:2004al}.
The ALMA observations of SgrA* will have a flux error under $0.05$~Jy, which is enough to reveal the predicted oscillations were they present on an observation night.
As our modeling indicates, the QPOs are absent when the magnetic field is weak due to destruction by magnetic field reversals.
 The future implementation of the Event Horizon Telescope may allow to measure the QPOs in the source size variations \citep{Doeleman:2009zn}.
The anticipated brighter state of SgrA* \citep{Moscibrodzka:2012av} after the cloud infall may alter our predictions: at a constant $\nu$
the flux increase is compensated by the lower fractional level of the QPOs due to the higher optical depth.

\section{ACKNOWLEDGEMENTS}\label{sec:acknowledgements}
The authors thank Chris Reynolds, Jim Moran, and the anonymous referee for comments.
This work was supported by NASA Hubble Fellowship grant HST-HF-51298.01 (RVS) and NSF/XSEDE resources provided by
NICS (Kraken/Nautilus) under the awards TG-AST080025N (JCM) and PHY120005 (RVS/JCM).
\bibliographystyle{apj}

\begin{thebibliography}{52}
\expandafter\ifx\csname natexlab\endcsname\relax\def\natexlab#1{#1}\fi

\bibitem[{{Abramowicz} {et~al.}(2003){Abramowicz}, {Karas}, {Kluzniak}, {Lee},
  \& {Rebusco}}]{Abramowicz:2003lu}
{Abramowicz}, M.~A., {Karas}, V., {Kluzniak}, W., {Lee}, W.~H., \& {Rebusco},
  P. 2003, \pasj, 55, 467

\bibitem[{{Abramowicz} \& {Klu{\'z}niak}(2001)}]{Abramowicz:2001ja}
{Abramowicz}, M.~A., \& {Klu{\'z}niak}, W. 2001, \aap, 374, L19

\bibitem[{Blandford \& Znajek(1977)}]{Blandford1977}
Blandford, R.~D., \& Znajek, R.~L. 1977, \mnras, 179, 433

\bibitem[{{Brown} {et~al.}(2004){Brown}, {Wild}, \&
  {Cunningham}}]{Brown:2004al}
{Brown}, R.~L., {Wild}, W., \& {Cunningham}, C. 2004, Advances in Space
  Research, 34, 555

\bibitem[{{Chan} {et~al.}(2009){Chan}, {Liu}, {Fryer}, {Psaltis}, {{\"O}zel},
  {Rockefeller}, \& {Melia}}]{Chan:2009ja}
{Chan}, C.-k., {Liu}, S., {Fryer}, C.~L., {Psaltis}, D., {{\"O}zel}, F.,
  {Rockefeller}, G., \& {Melia}, F. 2009, \apj, 701, 521

\bibitem[{{Dexter} {et~al.}(2010){Dexter}, {Agol}, {Fragile}, \&
  {McKinney}}]{Dexter:2010lk}
{Dexter}, J., {Agol}, E., {Fragile}, P.~C., \& {McKinney}, J.~C. 2010, \apj,
  717, 1092

\bibitem[{{Do} {et~al.}(2009){Do}, {Ghez}, {Morris}, {Yelda}, {Meyer}, {Lu},
  {Hornstein}, \& {Matthews}}]{Do:2009lk}
{Do}, T., {Ghez}, A.~M., {Morris}, M.~R., {Yelda}, S., {Meyer}, L., {Lu},
  J.~R., {Hornstein}, S.~D., \& {Matthews}, K. 2009, \apj, 691, 1021

\bibitem[{{Doeleman} {et~al.}(2009){Doeleman}, {Fish}, {Broderick}, {Loeb}, \&
  {Rogers}}]{Doeleman:2009zn}
{Doeleman}, S.~S., {Fish}, V.~L., {Broderick}, A.~E., {Loeb}, A., \& {Rogers},
  A.~E.~E. 2009, \apj, 695, 59

\bibitem[{{Doeleman} {et~al.}(2008){Doeleman}, {Weintroub}, {Rogers},
  {Plambeck}, {Freund}, {Tilanus}, {Friberg}, {Ziurys}, {Moran}, {Corey},
  {Young}, {Smythe}, {Titus}, {Marrone}, {Cappallo}, {Bock}, {Bower},
  {Chamberlin}, {Davis}, {Krichbaum}, {Lamb}, {Maness}, {Niell}, {Roy},
  {Strittmatter}, {Werthimer}, {Whitney}, \& {Woody}}]{Doeleman:2008af}
{Doeleman}, S.~S., {et~al.} 2008, \nat, 455, 78

\bibitem[{{Dolence} {et~al.}(2009){Dolence}, {Gammie}, {Mo{\'s}cibrodzka}, \&
  {Leung}}]{Dolence:2009wb}
{Dolence}, J.~C., {Gammie}, C.~F., {Mo{\'s}cibrodzka}, M., \& {Leung}, P.~K.
  2009, \apjs, 184, 387

\bibitem[{{Dolence} {et~al.}(2012){Dolence}, {Gammie}, {Shiokawa}, \&
  {Noble}}]{Dolence:2012xs}
{Dolence}, J.~C., {Gammie}, C.~F., {Shiokawa}, H., \& {Noble}, S.~C. 2012,
  \apjl, 746, L10

\bibitem[{{Drappeau} {et~al.}(2013){Drappeau}, {Dibi}, {Dexter}, {Markoff}, \&
  {Fragile}}]{Drappeau:2012dq}
{Drappeau}, S., {Dibi}, S., {Dexter}, J., {Markoff}, S., \& {Fragile}, P.~C.
  2013, \mnras, 431, 2872

\bibitem[{{Falanga} {et~al.}(2007){Falanga}, {Melia}, {Tagger}, {Goldwurm}, \&
  {B{\'e}langer}}]{Falanga:2007uj}
{Falanga}, M., {Melia}, F., {Tagger}, M., {Goldwurm}, A., \& {B{\'e}langer}, G.
  2007, \apjl, 662, L15

\bibitem[{{Falcke} {et~al.}(2004){Falcke}, {K{\"o}rding}, \&
  {Markoff}}]{Falcke:2004oq}
{Falcke}, H., {K{\"o}rding}, E., \& {Markoff}, S. 2004, \aap, 414, 895

\bibitem[{{Gammie} {et~al.}(2003){Gammie}, {McKinney}, \&
  {T{\'o}th}}]{Gammie:2003in}
{Gammie}, C.~F., {McKinney}, J.~C., \& {T{\'o}th}, G. 2003, \apj, 589, 444

\bibitem[{{Genzel} {et~al.}(2003){Genzel}, {Sch{\"o}del}, {Ott}, {Eckart},
  {Alexander}, {Lacombe}, {Rouan}, \& {Aschenbach}}]{Genzel:2003aj}
{Genzel}, R., {Sch{\"o}del}, R., {Ott}, T., {Eckart}, A., {Alexander}, T.,
  {Lacombe}, F., {Rouan}, D., \& {Aschenbach}, B. 2003, \nat, 425, 934

\bibitem[{{Ghez} {et~al.}(2008){Ghez}, {Salim}, {Weinberg}, {Lu}, {Do}, {Dunn},
  {Matthews}, {Morris}, {Yelda}, {Becklin}, {Kremenek}, {Milosavljevic}, \&
  {Naiman}}]{Ghez2008}
{Ghez}, A.~M., {et~al.} 2008, \apj, 689, 1044

\bibitem[{{Gierli{\'n}ski} {et~al.}(2008){Gierli{\'n}ski}, {Middleton}, {Ward},
  \& {Done}}]{Gierlinski:2008wj}
{Gierli{\'n}ski}, M., {Middleton}, M., {Ward}, M., \& {Done}, C. 2008, \nat,
  455, 369

\bibitem[{{Gillessen} {et~al.}(2009){Gillessen}, {Eisenhauer}, {Trippe},
  {Alexander}, {Genzel}, {Martins}, \& {Ott}}]{Gillessen:2009oo}
{Gillessen}, S., {Eisenhauer}, F., {Trippe}, S., {Alexander}, T., {Genzel}, R.,
  {Martins}, F., \& {Ott}, T. 2009, \apj, 692, 1075

\bibitem[{{Hamaus} {et~al.}(2009){Hamaus}, {Paumard}, {M{\"u}ller},
  {Gillessen}, {Eisenhauer}, {Trippe}, \& {Genzel}}]{Hamaus:2009fs}
{Hamaus}, N., {Paumard}, T., {M{\"u}ller}, T., {Gillessen}, S., {Eisenhauer},
  F., {Trippe}, S., \& {Genzel}, R. 2009, \apj, 692, 902

\bibitem[{{Henisey} {et~al.}(2012){Henisey}, {Blaes}, \&
  {Fragile}}]{Henisey:2012la}
{Henisey}, K.~B., {Blaes}, O.~M., \& {Fragile}, P.~C. 2012, \apj, 761, 18

\bibitem[{{Henisey} {et~al.}(2009){Henisey}, {Blaes}, {Fragile}, \&
  {Ferreira}}]{Henisey:2009gk}
{Henisey}, K.~B., {Blaes}, O.~M., {Fragile}, P.~C., \& {Ferreira}, B.~T. 2009,
  \apj, 706, 705

\bibitem[{{Kato}(2001)}]{Kato:2001aj}
{Kato}, S. 2001, \pasj, 53, 1

\bibitem[{{Kato}(2004)}]{Kato:2004fw}
{Kato}, Y. 2004, \pasj, 56, 931

\bibitem[{{Li} \& {Narayan}(2004)}]{Li:2004qp}
{Li}, L.-X., \& {Narayan}, R. 2004, \apj, 601, 414

\bibitem[{{Marrone}(2006)}]{Marrone:2006phd}
{Marrone}, D.~P. 2006, PhD thesis, Harvard University

\bibitem[{{Marrone} {et~al.}(2007){Marrone}, {Moran}, {Zhao}, \&
  {Rao}}]{Marrone:2007}
{Marrone}, D.~P., {Moran}, J.~M., {Zhao}, J.-H., \& {Rao}, R. 2007, \apjl, 654,
  L57

\bibitem[{{Marrone} {et~al.}(2008){Marrone}, {Baganoff}, {Morris}, {Moran},
  {Ghez}, {Hornstein}, {Dowell}, {Mu{\~n}oz}, {Bautz}, {Ricker}, {Brandt},
  {Garmire}, {Lu}, {Matthews}, {Zhao}, {Rao}, \& {Bower}}]{Marrone:2008ep}
{Marrone}, D.~P., {et~al.} 2008, \apj, 682, 373

\bibitem[{{Mauerhan} {et~al.}(2005){Mauerhan}, {Morris}, {Walter}, \&
  {Baganoff}}]{Mauerhan:2005rh}
{Mauerhan}, J.~C., {Morris}, M., {Walter}, F., \& {Baganoff}, F.~K. 2005,
  \apjl, 623, L25

\bibitem[{{McKinney} {et~al.}(2012){McKinney}, {Tchekhovskoy}, \&
  {Blandford}}]{McKinney2012}
{McKinney}, J.~C., {Tchekhovskoy}, A., \& {Blandford}, R.~D. 2012, \mnras, 423,
  3083

\bibitem[{{Miyoshi} {et~al.}(2011){Miyoshi}, {Shen}, {Oyama}, {Takahashi}, \&
  {Kato}}]{Miyoshi:2011ah}
{Miyoshi}, M., {Shen}, Z.-Q., {Oyama}, T., {Takahashi}, R., \& {Kato}, Y. 2011,
  \pasj, 63, 1093

\bibitem[{{Mo{\'s}cibrodzka} {et~al.}(2009){Mo{\'s}cibrodzka}, {Gammie},
  {Dolence}, {Shiokawa}, \& {Leung}}]{Moscibrodzka:2009}
{Mo{\'s}cibrodzka}, M., {Gammie}, C.~F., {Dolence}, J.~C., {Shiokawa}, H., \&
  {Leung}, P.~K. 2009, \apj, 706, 497

\bibitem[{{Mo{\'s}cibrodzka} {et~al.}(2012){Mo{\'s}cibrodzka}, {Shiokawa},
  {Gammie}, \& {Dolence}}]{Moscibrodzka:2012av}
{Mo{\'s}cibrodzka}, M., {Shiokawa}, H., {Gammie}, C.~F., \& {Dolence}, J.~C.
  2012, \apjl, 752, L1

\bibitem[{{Nowak} \& {Wagoner}(1991)}]{Nowak:1991jp}
{Nowak}, M.~A., \& {Wagoner}, R.~V. 1991, \apj, 378, 656

\bibitem[{{Papadakis} \& {Lawrence}(1993)}]{Papadakis93}
{Papadakis}, I.~E., \& {Lawrence}, A. 1993, \mnras, 261, 612

\bibitem[{{Penna} {et~al.}(2010){Penna}, {McKinney}, {Narayan}, {Tchekhovskoy},
  {Shafee}, \& {McClintock}}]{Penna:2010dj}
{Penna}, R.~F., {McKinney}, J.~C., {Narayan}, R., {Tchekhovskoy}, A., {Shafee},
  R., \& {McClintock}, J.~E. 2010, \mnras, 408, 752

\bibitem[{{Reis} {et~al.}(2012){Reis}, {Miller}, {Reynolds}, {G{\"u}ltekin},
  {Maitra}, {King}, \& {Strohmayer}}]{Reis:2012sc}
{Reis}, R.~C., {Miller}, J.~M., {Reynolds}, M.~T., {G{\"u}ltekin}, K.,
  {Maitra}, D., {King}, A.~L., \& {Strohmayer}, T.~E. 2012, Science, 337, 949

\bibitem[{{Remillard} \& {McClintock}(2006)}]{Remillard:2006}
{Remillard}, R.~A., \& {McClintock}, J.~E. 2006, Ann. Rev. Astron. Astr., 44,
  49

\bibitem[{{Schnittman} {et~al.}(2006){Schnittman}, {Krolik}, \&
  {Hawley}}]{Schnittman:2006ia}
{Schnittman}, J.~D., {Krolik}, J.~H., \& {Hawley}, J.~F. 2006, \apj, 651, 1031

\bibitem[{{Shcherbakov} \& {Baganoff}(2010)}]{Shcherbakov:2010cond}
{Shcherbakov}, R.~V., \& {Baganoff}, F.~K. 2010, \apj, 716, 504

\bibitem[{{Shcherbakov} \& {Huang}(2011)}]{Shcherbakov:2011inter}
{Shcherbakov}, R.~V., \& {Huang}, L. 2011, \mnras, 410, 1052

\bibitem[{{Shcherbakov} {et~al.}(2012){Shcherbakov}, {Penna}, \&
  {McKinney}}]{Shcherbakov:2012appl}
{Shcherbakov}, R.~V., {Penna}, R.~F., \& {McKinney}, J.~C. 2012, \apj, 755, 133

\bibitem[{{Shiokawa} {et~al.}(2012){Shiokawa}, {Dolence}, {Gammie}, \&
  {Noble}}]{Shiokawa:2012iq}
{Shiokawa}, H., {Dolence}, J.~C., {Gammie}, C.~F., \& {Noble}, S.~C. 2012,
  \apj, 744, 187

\bibitem[{{Stone} \& {Gardiner}(2007)}]{Stone:2007qw}
{Stone}, J.~M., \& {Gardiner}, T. 2007, \apj, 671, 1726

\bibitem[{{Tagger} \& {Melia}(2006)}]{Tagger:2006jw}
{Tagger}, M., \& {Melia}, F. 2006, \apjl, 636, L33

\bibitem[{{Timmer} \& {Koenig}(1995)}]{Timmer:1995wa}
{Timmer}, J., \& {Koenig}, M. 1995, \aap, 300, 707

\bibitem[{{Trippe} {et~al.}(2007){Trippe}, {Paumard}, {Ott}, {Gillessen},
  {Eisenhauer}, {Martins}, \& {Genzel}}]{Trippe:2007oa}
{Trippe}, S., {Paumard}, T., {Ott}, T., {Gillessen}, S., {Eisenhauer}, F.,
  {Martins}, F., \& {Genzel}, R. 2007, \mnras, 375, 764

\bibitem[{{van der Klis}(2000)}]{Klis:2000he}
{van der Klis}, M. 2000, \araa, 38, 717

\bibitem[{{Wagoner}(2008)}]{Wagoner:2008pq}
{Wagoner}, R.~V. 2008, New Astronomy Reviews, 51, 828

\bibitem[{{Yusef-Zadeh} {et~al.}(2011){Yusef-Zadeh}, {Wardle}, {Miller-Jones},
  {Roberts}, {Grosso}, \& {Porquet}}]{Yusef-Zadeh:2011os}
{Yusef-Zadeh}, F., {Wardle}, M., {Miller-Jones}, J.~C.~A., {Roberts}, D.~A.,
  {Grosso}, N., \& {Porquet}, D. 2011, \apj, 729, 44

\end{thebibliography}

\end{document}